# Coupled macro spin model with two variables for polarity reversals in the Earth and the Sun

Ariyoshi Kunitomo[1], Akika Nakamichi[2*], Tetsuya Hara[3]

[1] *Sendai Astronomical Observatory, Sendai 989-3123, Japan*
[2] *Institute of General Education, Kyoto Sangyo University, Kyoto 603-8555, Japan*
[3] *Department of Physics, Kyoto Sangyo University, Kyoto 603-8555, Japan*
*E-mail: nakamich3@gmail.com

## Abstract

The structure of geomagnetism is very complex, and there are still some problems left in magnetohydrodynamics (MHD) simulations. Recently, the macro spin model has been suggested. This is the idea that geomagnetism is described by interaction with many local dynamo elements (macro-spins). This model can reproduce some features of geomagnetism by solving equations of motion with only one variable. In this paper, we study this model to make more general by adding one variable. In this result, our model becomes possible for several things which are not treated in the previous study, for example, migration of the North (or South) Magnetic Pole, comparison with observed data of magnetic field distributions expressed in two directions, etc. Moreover, as a result of application to the sun, we could reproduce periodic polarity reversals and the power index of the power spectrum, etc. In addition, we investigate the statistical properties of the pole migration.

## I. INTRODUCTION

Geomagnetism has some unsolved problems. One of them is polarity reversal and it is known that 332 reversals are observed in the last $1.6 \times 10^8$ years (Fig. 1) [1,2]. A simple disk model [3] can explain polarity reversal, however, actual geomagnetism is very complex. Today, geodynamo is studied by numerical simulations [4-6], and some dynamo experiments [7]. They reproduced partly, some polarity reversals, etc. However, there are still some problems that the number of reversals is less than observation and the values of the parameters in numerical simulation are far from the real values [4]. Moreover, reasons which are the cause of polarity reversals are not clearly known [6,8].

Recently, the macro spin model for polarity reversals has been suggested [9]. This model is based on the idea that electric current winds for vortex structures generated by convective structures in the Earth. This electric current forms magnetic field that has macro direction, which we call a macro-spin. The behavior of whole dynamo mechanism is described by interactions with these macro-spins. Although this model can reproduce many features of geomagnetism, this model has only one variable $\theta$ which represents angle with respect to the rotational axis. In addition, this model is applicable to the Sun, and other celestial objects.

In this paper, we suggest that macro spin model with two variables, being added a new variable $\phi$ which represents longitude. In the results of numerical calculations with two variables, we can reproduce both polarity reversals and migration of the North (or South) Magnetic Pole. This means that macro spin model becomes more general than previous one [9].

## II. LONG-RANGE COUPLED SPIN (LCS) MODEL WITH TWO VARIABLES

The macro spin model has two types. One is the short-range coupled spin (SCS) model where macro-spins interact with only neighboring spins [10]. In the SCS model, friction and random force terms are necessary for reversal (Langevin-type equation). The other type is long-range coupled spin (LCS) model where spins interact with all other spins. In this paper, we adopt LCS model. We introduce the longitudinal angle $\phi$ for the macro spin model. By spherical coordinate parameters, the spin $\vec{S}_i$ is described $\vec{S}_i = (\sin\theta_i \cos\phi_i, \sin\theta_i \sin\phi_i, \cos\theta_i)$, ($i = 1, 2, \cdots, N$), where $N$ is the number of spins. Thus, the kinetic energy $T$ of the system is

$$T \equiv \frac{1}{2}\sum_{i=1}^{N}\left(\frac{d\vec{S}_i}{dt}\right)^2 = \frac{1}{2}\sum_{i=1}^{N}(\dot{\theta}_i^{\,2} + \dot{\phi}_i^{\,2}\sin^2\theta_i). \tag{1}$$

We adopt this $T$ as kinetic energy of macro spin model with two variables for each spin.

Using the same equation for the potential energy $V$ in previous study [9],

$$V = \mu \sum_{i=1}^{N}(\vec{\Omega}\cdot\vec{S}_i)^2 + \frac{\lambda}{2N}\sum_{i<j}^{N}(\vec{S}_i\cdot\vec{S}_j), \tag{2}$$

we get the Lagrangian $L = T - V$ for this system.

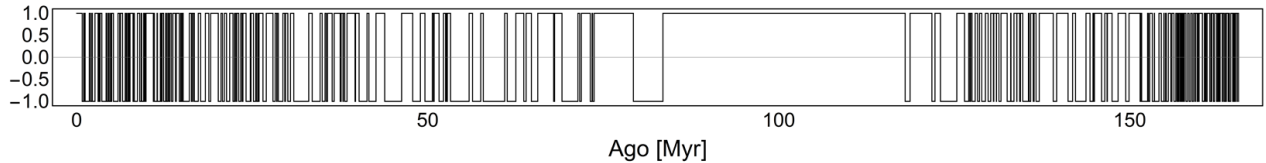

FIG. 1: Geomagnetic polarity reversal record by observation in the last $1.6 \times 10^8$ years. This record has 332 times polarity reversal. $+1.0$ represents the same direction and $-1.0$ represents opposite direction at the present day [1].

As a result we get equations of motion for $\theta_i$ and $\phi_i$ from these Lagrangian;

$$\frac{d}{dt}\left(\frac{\partial L}{\partial \dot{\theta}_i}\right) = \frac{\partial L}{\partial \theta_i}, \tag{3}$$

and

$$\frac{d}{dt}\left(\frac{\partial L}{\partial \dot{\phi}_i}\right) = \frac{\partial L}{\partial \phi_i}. \tag{4}$$

In the LCS model, friction and random force terms are not necessary for reversal. Recently, however, the paper that investigated the Langevin-type equation in the LCS model has been suggested [11]. We solve equations (3) and (4), and study the behavior.

We define "magnetization" $M(t)$:

$$M(t) \equiv \frac{1}{N}\sum_{i=1}^{N} \vec{\Omega} \cdot \vec{S}_i = \frac{1}{N}\sum_{i=1}^{N} \cos\theta_i, \tag{5}$$

as an indicator of polarity. $M(t)$ is an average of the projection of spins onto the rotational axis $\vec{\Omega}$, where $\vec{\Omega} = (0,0,1)$. As the indicator $M(t)$ represents the direction of polarity, the inversion of the sign of $M(t)$ corresponds to the polarity reversal.

## III.  NUMERICAL RESULTS AND COMPARISON WITH OBSERVED DATA

We assume initial conditions that all macro-spins latitude angle $\theta_i$ is within a range of $\pm \pi/4$ around the pole to investigate whether flipping from the state where all macro-spins direction are aligned is possible. When we set the parameters are $N = 9$, $\mu = -11$, and $\lambda = -18.4$, we obtained 267 reversals in $3 \times 10^5$ calculation time. These parameter values are determined to reproduce two observed geomagnetic data: the polarity-reversal number in the last $1.6 \times 10^8$ years and pole migration length in the last $1.6 \times 10^3$ years (from 200 to 1800 AD). These numerical results are calculated by using Mathematica 10 platform. In Figs. 2 and 3, the time series and the power spectrum of $M(t)$ are presented. In Fig. 4, the power spectrum of observed data is shown [1,2]. Comparing with these results and observed data, we obtain following results.

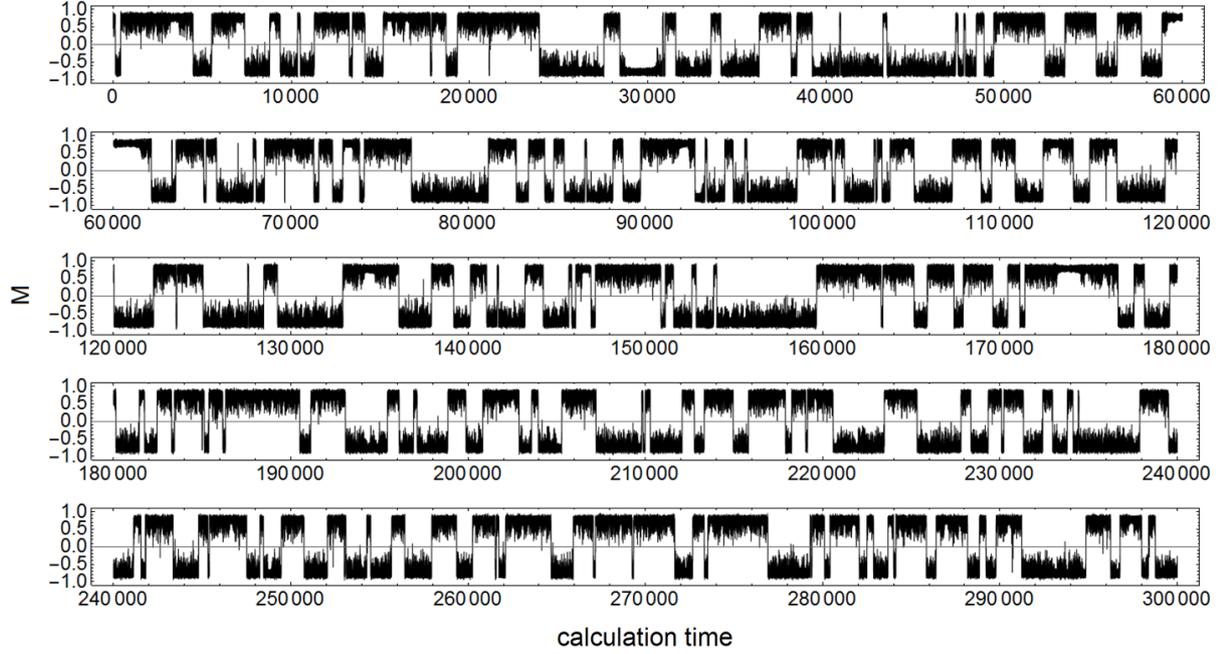

FIG. 2: The time series of $M(t)$ in the macro spin model of two variables with the parameters $N = 9$, $\mu = -11$ and $\lambda = -18.4$. In this case, there appears 267 times reversal in $3 \times 10^5$ calculation time. We see a lot of random and rapid reversals.

(i) Unit calculation time: $4.3 \times 10^2$ years.
(ii) The average time of polarity flipping: $2 \times 10^3$ years (observation: $(2 - 3) \times 10^3$ years).
(iii) Variety of chron: $(0.06 - 2.5) \times 10^6$ years (observation: $(0.1 - 6) \times 10^6$ years).
(iv) The power index of the power spectrum of $M(t)$: $-0.15$ and $-1.96$ (observation: $-0.76$ and $-1.77$, respectively).
(v) The mean pole migration length every $50$ years when assuming that the iron fluid core as a unit sphere: $0.1$ (observation: $0.098$ [12]).

Thus, the calculation of the adopted parameters in our model can reproduce some features of real geomagnetic behaviors. Moreover, we obtain some features of polarity behaviors since this model contains two variables. In Fig. 5, a snapshot of polarity flip is shown with times where spins move from the top to the bottom together. At the moment of polarity flip, dipole magnetic field is expected to be weak. A migration of pole is represented in Fig. 6 (a) and (b), where the thick line is the time series of trajectory. These trajectories are based on observed paleomagnetic data of migration of the North Magnetic Pole (Fig. 6 (a)) [12] and drawn by the end point of magnetization vector $M(t)$ in the numerical calculation in our model (Fig. 6 (b)). The circular lines in Fig.6 represent latitude lines every ten degrees viewed from the north. The observed data is similar to the simulation results that show random migration of pole around the North Pole and mean migration length every $50$ years.

There are many paleomagnetic observations which show preferential routes for the migration from one pole to another [13]. This model shows a kind of "isotropy" as presented in Fig. 6 (b), so it does not reproduce one of the main open problems of paleomagnetic observations.

For superchron is discussed in *section 6*.

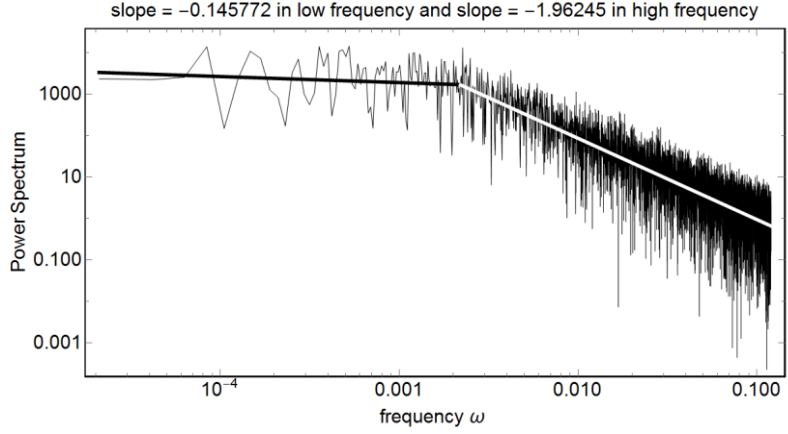

FIG. 3: The power spectrum of $M(t)$ in the macro spin model with two variables. Two power laws indicate $-0.15$ in low-frequency region and $-1.96$ in high-frequency region.

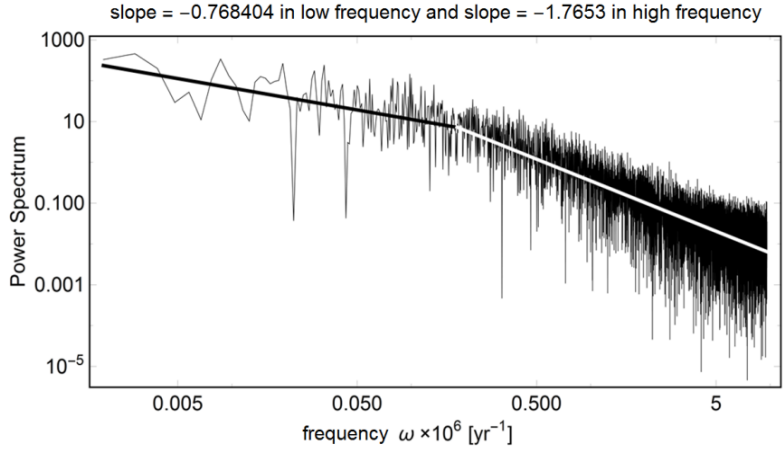

FIG. 4: The power spectrum of the observed data [1,2]. Two power laws indicate $-0.77$ in low-frequency region and $-1.8$ in high- frequency region.

## IV.  APPLICATION TO THE SUN

The previous study [9] has shown that macro spin model is applicable not only to geomagnetism but also to solar magnetism. We use a scaling law found in previous study, where 10 celestial objects for the mass range over 8-digits are located on a straight line which draws a graph for $d/(m^{1/2}R^3\ (2\rho_0\Omega/\sigma)^{1/2})$ against $m$, where $d$, $m$, $R$, $\rho_0$, and $\sigma$ are magnetic dipole moment, mass of celestial object, radius of celestial object, mass density and electric conductivity of outer core respectively. This scaling laws and analogy from MHD equations lead to proportional relation between spin number and mass of celestial object;

$$N\gamma^2 \propto m^{1/2}. \qquad (6)$$

If the parameter $\gamma$, which represents the ratio of radius of the Taylor cell and the radius of the iron core, is constant for various objects, the number of spins in the Sun is determined to $N_{\text{sun}} \approx 5 \times 10^3$, under using the number of spins in the Earth $N = 9$. Therefore, we investigate whether the macro spin model of two variables can describe solar magnetism by using a large spin number.

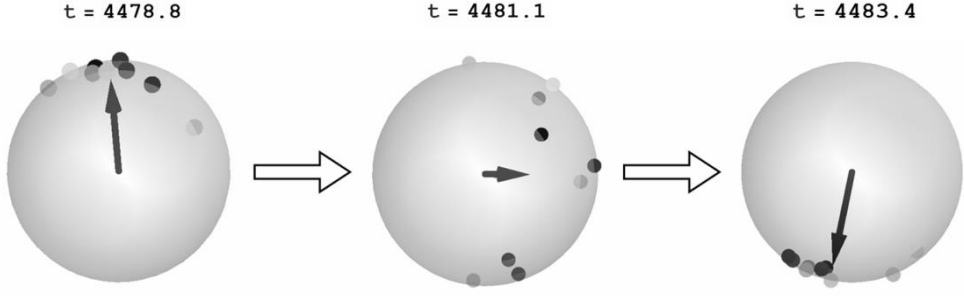

FIG. 5: A snapshot at the moment of polarity flipping. The black arrow from the center of the sphere represents "magnetization" $M(t)$. Almost all spins are moving from the top to the bottom together.

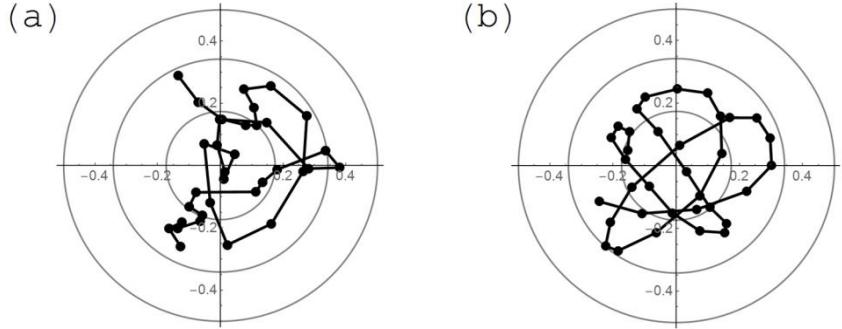

FIG. 6: (a) A trajectory of migration of the North Magnetic Pole that based on observed paleomagnetic data from 200 to 1800 AD. (b) A typical trajectory of migration of pole in our model ($t = 3643.47 - 3647.28$). The thick line is time series of a trajectory drawn by the end point of magnetization vector $M(t)$. We see the random magnetic pole migration around the north pole in (b) is similar to observation data in (a).

We assume the initial conditions for the Sun that all macro-spins latitude angle $\theta_i$ are within a range of $\pm \pi/2$ around the pole.[1] For the case of spin number $N = 300$, the parameters $\mu = -0.3$ and $\lambda = -2$, we obtain time series of $M(t)$ shown in Fig. 7 and power spectrum of $M(t)$ shown in Fig. 8. In Fig. 7, periodic pattern of the feature of the solar magnetism is presented. We can expect that short period corresponds to approx 11 years period. We could also expect that places of weak magnetic field intensity become weak, e. g. around the $t \approx 170$, correspond to the time when solar magnetism activity became weak. Moreover, the peak of corresponding 11 years period exists around $\omega = 2$ in Fig. 8 and the power spectrum behaves close to $1/f$ noise because the power index is $-1$ [14].

From the above results, the macro spin model with two variables can reproduce the features of solar magnetism behavior, 11 years period and $1/f$ noise of power spectrum. If we take the large number $N$ ($> 100$), the periodic features could be noticed.

---

[1] The reason for assuming the initial conditions, it is considered to the macro-spins directed wider range than the case of the Earth because of the number of spins is larger in the Sun.

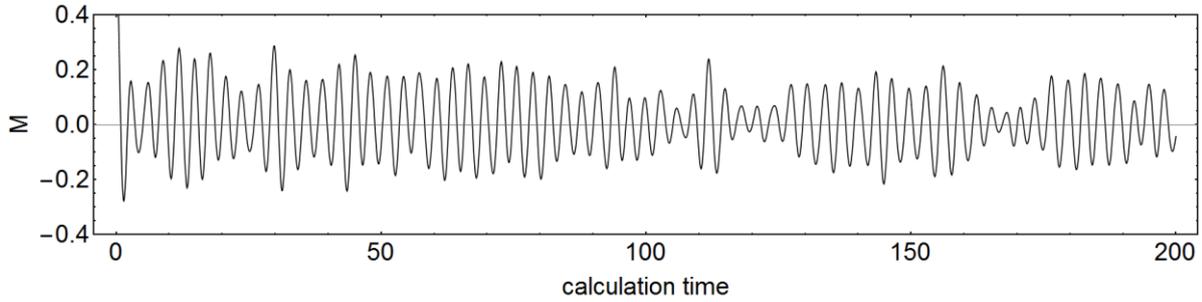

FIG. 7: The time series of $M(t)$ in the macro spin model with two variables for the Sun. The parameters are $N = 300$, $\mu = -0.3$, $\lambda = -2$. The periodic behavior of reversal is different from the behavior in geomagnetism. We can expect that short reversing period corresponds to 11 years period, and the places of weak magnetic field intensity correspond to the time when solar magnetic activity is weak.

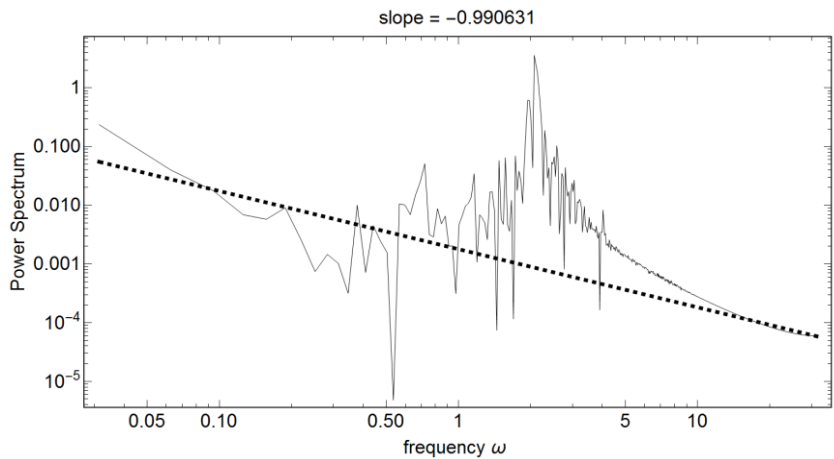

FIG. 8: The power spectrum of $M(t)$ in the macro spin model with two variables for the Sun (solid line) and the power index (dashed line). The peak around $\omega = 2$ corresponds to 11 years period and the power index is $-1$ similar to $1/f$ noise.

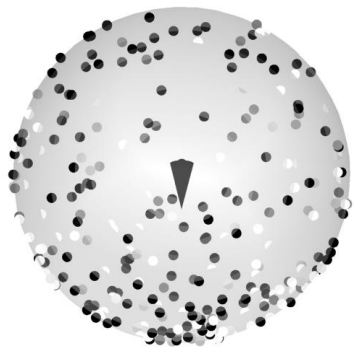

FIG. 9: A snapshot of the Sun in the macro spin model with two variables. The arrow from the center of the sphere (inverted triangle) represents "magnetization" $M(t)$. The polarity direction is downward in this figure while there are a lot of spins which point to opposite direction of the polarity. Therefore if the upper side of this figure is north, there does not exist only dominated negative sign of magnetic fields, but also positive sign in the north polar region.

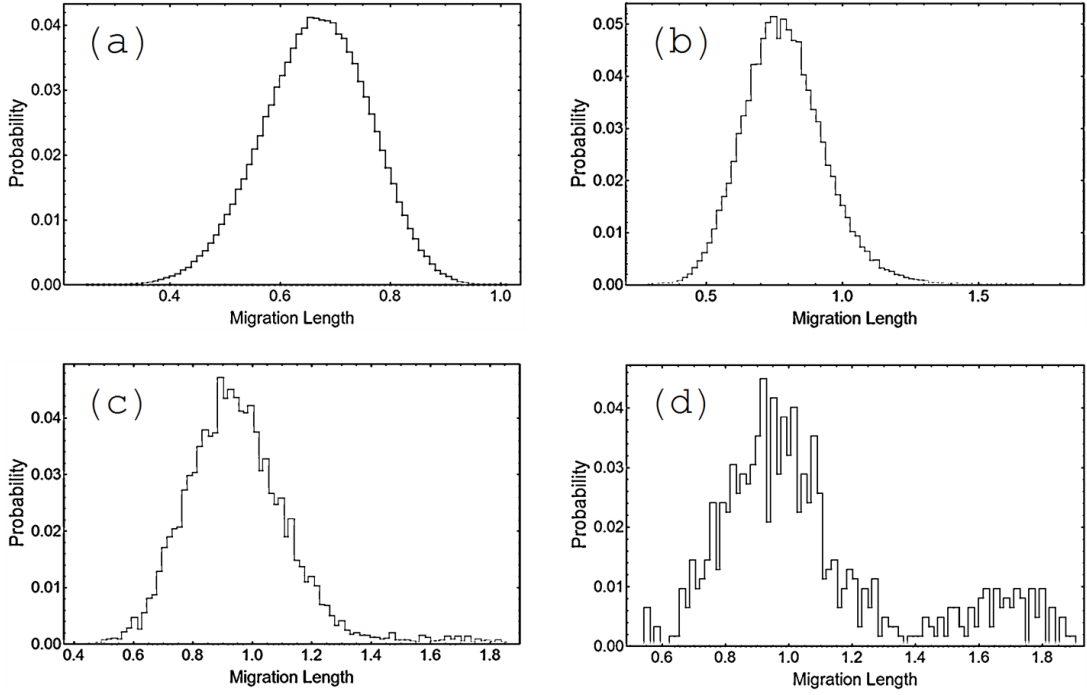

FIG. 10: The histograms of probability distribution in our model. (a) $\Delta t = 0.24$ (about 100 years). The fit distribution function for this histogram is similar to normal distribution. (b) $\Delta t = 2.4$ (about $10^3$ years). (c) $\Delta t = 24$ (about $10^4$ years). This histogram has a long tail in the right side. (d) $\Delta t = 240$ (about $10^5$ years). Another peak appears on the right side of the histogram around 1.7 at migration length by the effect of polarity reversals.

In Fig. 9, a snapshot at a certain moment of all spins for the Sun is shown. There still exist a lot of macro-spins pointing to opposite direction to the polarity in the figure. In fact, both positive and negative magnetic field in the north polar region of the Sun was observed [15]. The macro spin model with one variable is not able to compare with such observed data expressed in spherical distribution, therefore this comparison is one of the advantages of the macro spin model with two variables.

## V. Statistical properties of pole migration

### A. Straight-line distance

We investigate the migration length of the pole by numerical simulation to examine the form of distribution in our model.

In Fig. 10, the histograms of probability distribution in our model are shown. The histogram represents event probability of the migration length in $\Delta t$ for all time steps number. The value of migration length is the straight-line distance between two points of the end point of magnetization vector $M(t)$ at the time $t_n$ and $t_{n-1}$, where $t_n = t_{n-1} + \Delta t$, $n$ is time step number, and $\Delta t$ is constant each histogram. In this case, the fit distribution function for the histogram is similar to normal distribution when $\Delta t$ is small, and shift to the form that has a tail on the right side when $\Delta t$ becomes large, because large $\Delta t$ contains polarity reversals. In Fig. 10 (d), the other peak appears on the right side of the histogram around 1.7 migration length by the effect of polarity reversals.

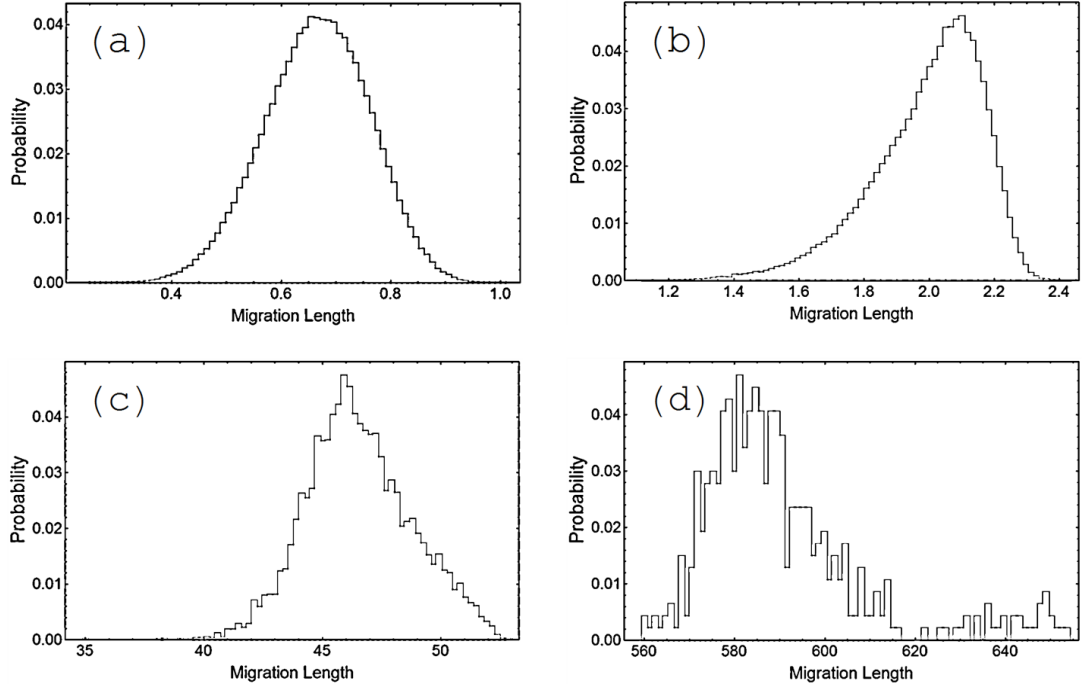

FIG. 11: The histograms of the probability distribution for total distance in our model. (a) $\Delta t = 0.24$ (about 100 years). The fit distribution function for this histogram is similar to normal distribution. (b) $\Delta t = 0.72$ (about 300 years). (c) $\Delta t = 16.8$ (about $7 \times 10^3$ years). This histogram has a long tail in the left side. (d) $\Delta t = 216$ (about $9 \times 10^4$ years). This histogram has a long tail in the right side.

From the results of Fig. 10, we can see the histogram of migration length becomes heavy (long)-tailed distribution. However, the information is lost that path between two points is lacked in this case. Therefore, we will investigate the migration length of total moving distance in next section.

### B. Total distance

In Fig. 11, the histograms of the probability distribution for total distance in our model are shown. In this case, the fit distribution function for the histogram is similar to normal distribution when $\Delta t$ is small too, however, Fig. 11 (b) becomes the form that has a tail on the left side when $\Delta t$ becomes larger. And Fig. 11 (c) becomes similar to normal distribution again. Finally, Fig. 11 (d) becomes heavy (long)-tailed distribution that is similar to Fig. 10 (d).

From these result, we reproduce one similar to heavy (long)-tailed distribution when $\Delta t$ is about several $10^5$ years scale, however, the shift in the location of the tail in histogram is not constant.

Therefore, we find that the migration length of the pole in the Earth does not follow one unique distribution function.

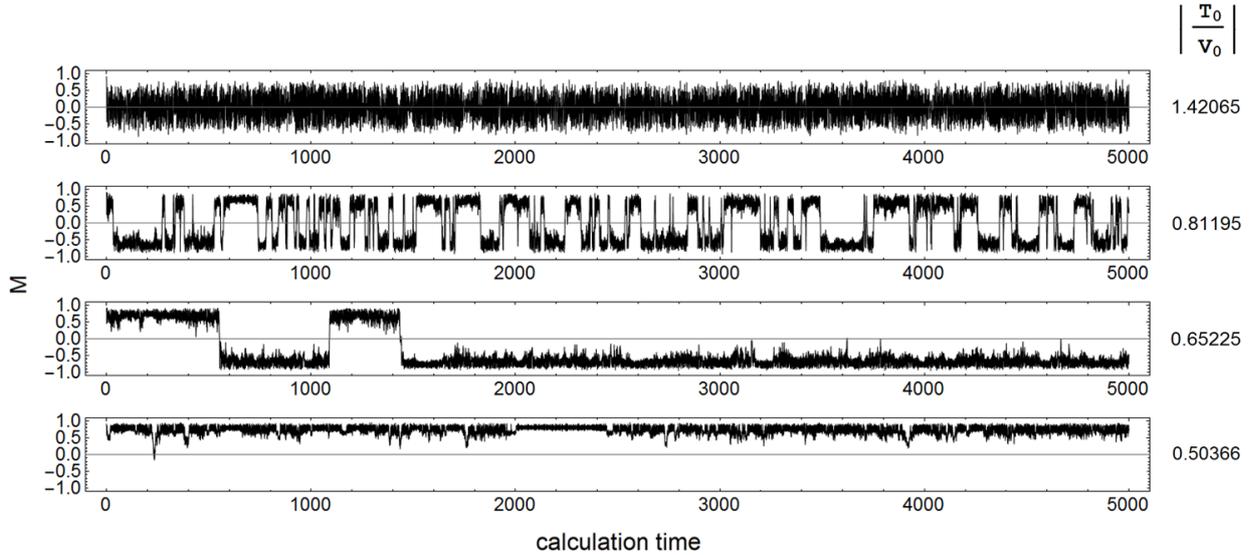

FIG. 12: The time series of $M(t)$ in the macro spin model with two variables comparing the value of $\mu$ under constant $\lambda = -18.4$. The value of $\mu$ is $0, -7, -11, -17$ from top to bottom. The values located on the right side of the graph are the absolute value of the ratio of initial kinetic energy to potential energy $|T_0/V_0|$. We see that the number of polarity reversals is decreasing as deepening the potential energy.

## VI. CONCLUSIONS AND DISCUSSIONS

We summarize the results of macro spin model with two variables in this paper. At first, we could reproduce the features of the geomagnetic and solar magnetic behaviors; power spectrum, the average time, and randomness of polarity reversals. For large $N$ ($> 100$) in solar case, it could be noticed the periodical feature. In low-frequency region, the power index of the Earth in our model is different from the observed data. However, this mismatch is not so important because a number of the data point is small in this region. In addition, a lack of superchron might be one of the causes of this mismatch. A superchron is the very long period of fixed polarity, it's time is longer than $1 \times 10^7$ years [16]. In the second, we could reproduce polarity random migration and the mean migration length of the magnetic pole by the introduction of longitudinal angle. In the third, kinetic energy form of the macro spin model with two variables suggested in this paper is more simple form than the other previous study [17]. Thus, we think that our model is suitable for extracting some essences of polarity reversal. In the fourth, it becomes possible to compare with the observed data of the Sun expressed in spherical distribution. Moreover, the state of the result of numerical calculations is similar to observed data [15]. In the fifth, we investigate the statistical properties of the magnetic pole. In the result, we find that the migration length of the pole in the Earth does not follow one unique distribution function.

The relationship between the parameters and observed values remain unclear. LCS model has two parameters $\mu$ and $\lambda$ which represent the strength of interaction between spins and rotational axis, and the strength of interaction between each spin, respectively. To change these parameters corresponds to change the depth of potential energy. In Fig.12, the graphs are compared with changing the parameter $\mu$, under the constant $\lambda$. In Fig.13, the graphs are compared with changing the parameter $\lambda$, under the constant $\mu$. These graphs show that the number of polarity reversals is decreasing with the smaller absolute value of the ratio of initial kinetic energy to potential energy. According to these results, we see the polarity is difficult to flip in deep potential energy. We adopt the values of parameters in the numerical calculation which are consistent with the observed data by this tendency.

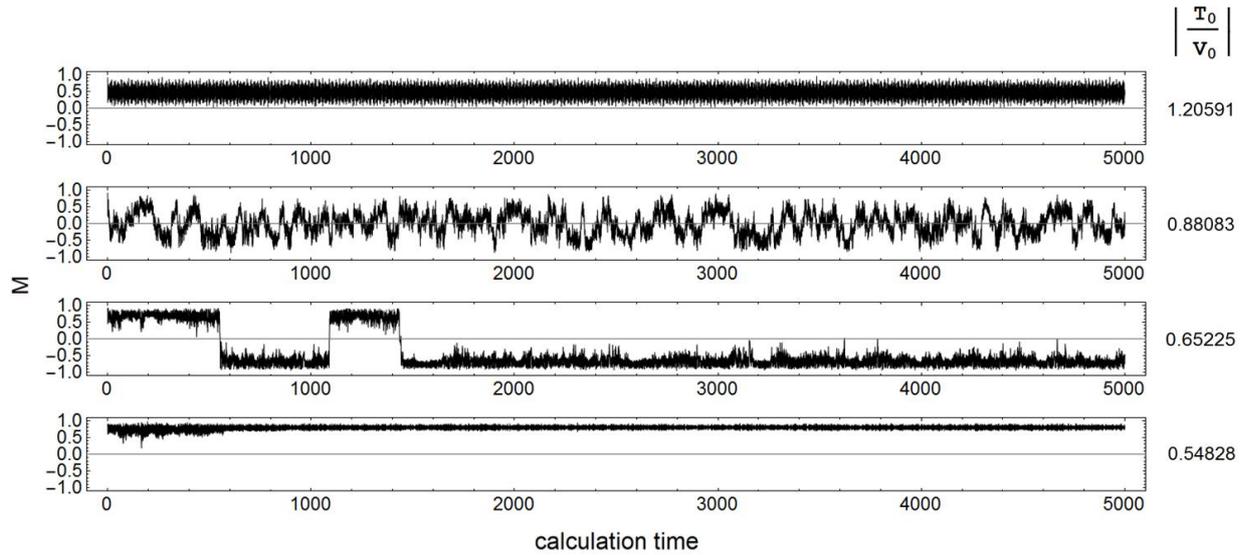

FIG. 13: The time series of $M(t)$ in the macro spin model with two variables comparing the value of $\lambda$ under constant $\mu = -11$. The value of $\lambda$ is $0, -8, -18.4, -26$ from top to bottom. The values located on the right side of the graph are the absolute value of the ratio of initial kinetic energy to potential energy $|T_0/V_0|$. We see that the number of polarity reversals is decreasing as deepening the potential energy.

If we could take the fine delicate values of $\lambda$ and $\mu$, it could be explained the superchron where there is almost no reversal. However, it seems to be difficult to implement the polarity reversals adequately before and after the superchron even if it is realized [16]. We would like to investigate the problem in the future.

### Acknowledgements

We would like to thank Y. Itoh for valuable discussions. This work was supported by KAKENHI, Grant-in-Aid for Scientific Research (C26400234).

### References

[1]  J. G. Ogg, Global Earth Physics, 240 (1995).

[2]  S. C. Cande, D. V. Kent, J. Geophys. Res., **100**, 6093–6095 (1995).

[3]  T. Rikitake, Proc. Cambridge Phil. Soc, **54**, 89 (1958).

[4]  A. Kageyama, T. Miyagoshi, T. Sato, Nature, **454**, 1106 (2008).

[5]  G. Glatzmaier, P. H. Roberts, Nature, **377**, 203 (1995).

[6] P. Olson, G. Glatzmaier, R. Coe, Earth. Planet. Sci. Lett., **304**, 168 (2011).

[7] M. Berhanu et al., EPL, **77**, 59001 (2007) [arXiv:physics/0701076 [physics.flu-dyn]].

[8] D. Gubbins, 2008, Earth Science, **452**, 13 (2008)

[9] A. Nakamichi, H. Mouri, D. Schmitt, A. Ferriz-Mas, J. Wicht, M. Morikawa, Mon. Not. R. Astron. Soc., **423**, 2977 (2012) [arXiv:1104.5093 [astro-ph.EP]].

[10] N. Mori, D. Schmitt, A. Ferriz-Mas, J. Wicht, H. Mouri, A. Nakamichi, M. Morikawa, Phys. Rev. E, **87**, 012108 (2013) [arXiv:1110.5062 [astro-ph.EP]].

[11] B. Duka, K. Peqini, A. De Santis, F. J. Pavón-Carrasco, Physics of the Earth and Planetary Interiors, **242**, 9 (2015).

[12] G. St-Onge, J. S. Stoner, Oceanography, **24(3)**, 42 (2011).

[13] T. Lay, Q. Williams, E. J. Garnero, Nature, **392**, 461 (1998)

[14] H. Fanchiotti, S. J. Sciutto, C. A. Garcia Canal, C. Hojvat, Fractals, **12**, 405 (2004) [arXiv:nlin/0403032 [nlin.AO]].

[15] H. Itoh, S. Tsuneta, D. Shiota, T. Tokumaru, K. Fujiki, The Astrophysical Journal, **719**, 131 (2010).

[16] D. R. Franco, W. P. Oliveira, F. V. Barbosa, D. Takahashi, C. F. P. Neto, I. M. C. Peixoto, Scientific Reports, **9**, 282 (2019).

[17] M. Yu. Reshetnyak, Izvestiya, Physics of the Solid Earth, **49**, 464 (2013)